\documentclass{nature}
\usepackage{amsmath,graphicx}
\usepackage{epsf}
\usepackage{epstopdf}
\usepackage{subfigure,epstopdf, textcomp}
\input{epsf}
\usepackage{epsfig}
\usepackage{color}
\usepackage{lscape}
\DeclareGraphicsExtensions{.jpg,.pdf,.png,.eps,.ps}

\def\red#1 {\textcolor{red}{#1}\ }

\def\ts     {\thinspace}   
\def\cone {\ifmmode{{\rm C}{\rm \small I}\,(^3P_1\to^3P_0)}\else{C\ts{\scriptsize I}\,{\small$(^3P_1\to^3P_0)$}}\fi}
\def\ci   {\ifmmode{{\rm C}{\rm \small I}}\else{C\ts {\scriptsize I}}\fi}
\def\arcsec {\mbox{$^{\prime\prime}$}}

\def\micron {$\mu$m}
\def\Caltech{1}
\def\Arizona{2}
\def\Dal{3}
\def\Cambridge{4}
\def\ESO{5}
\def\McGill{6}
\def\MPIfR{7}
\def\UPenn{8}
\def\UChicago{9}
\def\CfA{10}
\def\Harvard{11}
\def\KICPChicago{12}
\def\EFIChicago{13}
\def\PhysicsUChicago{14}
\def\JPL{15}
\def\Miss{16}
\def\AAUChicago{17}
\def\ANL{18}
\def\Davis{19}
\def\NRAO{20}
\def\Berkeley{21}
\def\UFlorida{22}
\def\UCL{23}
\def\Colorado{24}
\def\LBNL{25}
\def\UCLA{26}
\def\ATNF{27}
\def\Michigan{28}
\def\CaseWestern{29}
\def\AstroMichigan{30}
\def\Carnegie{31}
\def\STScI{32}
\def\SAIC{33}
\def\Yale{34}
\def\IAS{35}

\title{Dusty starburst galaxies in the early Universe as revealed by gravitational lensing}

\author{
J.~D.~Vieira$^{\Caltech}$,
 D.~P.~Marrone$^{\Arizona}$,   
 S.~C.~Chapman$^{\Dal,\Cambridge}$,
 C.~De~Breuck$^{\ESO}$,
 Y.~D.~Hezaveh$^{\McGill}$,
 A.~Wei\ss$^{\MPIfR}$,
 J.~E.~Aguirre$^{\UPenn}$,
 K.~A.~Aird$^{\UChicago}$,
 M.~Aravena$^{\ESO}$,
 M.~L.~N.~Ashby$^{\CfA}$,
 M.~Bayliss$^{\Harvard}$, 
B.~A.~Benson$^{\KICPChicago,\EFIChicago}$, 
 A.~D.~Biggs$^{\ESO}$,
L.~E.~Bleem$^{\KICPChicago,\PhysicsUChicago}$, 
 J.~J.~Bock$^{\Caltech,\JPL}$,
 M.~Bothwell$^{\Arizona}$,
 C.~M.~Bradford$^{\JPL}$,
M.~Brodwin$^{\Miss}$,
J.~E.~Carlstrom$^{\KICPChicago,\EFIChicago,\PhysicsUChicago,\AAUChicago,\ANL}$, 
C.~L.~Chang$^{\KICPChicago,\EFIChicago,\ANL}$, 
T.~M.~Crawford$^{\KICPChicago,\AAUChicago}$, 
A.~T.~Crites$^{\KICPChicago,\AAUChicago}$,
T.~de~Haan$^{\McGill}$, 
M.~A.~Dobbs$^{\McGill}$, 
E.~B.~Fomalont$^{\NRAO}$,
 C.~D.~Fassnacht$^{\Davis}$,
E.~M.~George$^{\Berkeley}$, 
M.~D.~Gladders$^{\KICPChicago,\AAUChicago}$, 
A.~H.~Gonzalez$^{\UFlorida}$, 
T.~R.~Greve$^{\UCL}$,	
B.~Gullberg$^{\ESO}$,	
N.~W.~Halverson$^{\Colorado}$, 
F.~W.~High$^{\KICPChicago,\AAUChicago}$, 
G.~P.~Holder$^{\McGill}$, 
W.~L.~Holzapfel$^{\Berkeley}$, 
S.~Hoover$^{\KICPChicago,\EFIChicago}$, 
J.~D.~Hrubes$^{\UChicago}$, 
T.~R.~Hunter$^{\NRAO}$,
R.~Keisler$^{\KICPChicago,\PhysicsUChicago}$, 
A.~T.~Lee$^{\Berkeley,\LBNL}$, 
E.~M.~Leitch$^{\KICPChicago,\AAUChicago}$, 
M.~Lueker$^{\Caltech}$, 
D.~Luong-Van$^{\UChicago}$, 
 M.~Malkan$^{\UCLA}$,
 V.~McIntyre$^{\ATNF}$,
J.~J.~McMahon$^{\KICPChicago,\EFIChicago,\Michigan}$, 
J.~Mehl$^{\KICPChicago,\AAUChicago}$, 
 K.~M.~Menten$^{\MPIfR}$,
S.~S.~Meyer$^{\KICPChicago,\EFIChicago,\PhysicsUChicago,\AAUChicago}$, 
L.~M.~Mocanu$^{\KICPChicago,\AAUChicago}$,
 E.~J.~Murphy$^{\Carnegie}$,
T.~Natoli,$^{\KICPChicago,\PhysicsUChicago}$, 
S.~Padin$^{\Caltech,\KICPChicago,\AAUChicago}$, 
T.~Plagge$^{\KICPChicago,\AAUChicago}$, 
C.~L.~Reichardt$^{\Berkeley}$, 
A.~Rest$^{\STScI}$, 
J.~Ruel$^{\Harvard}$, 
J.~E.~Ruhl$^{\CaseWestern}$, 
 K.~Sharon$^{\KICPChicago,\AAUChicago,\AstroMichigan}$,
K.~K.~Schaffer$^{\KICPChicago,\SAIC}$, 
L.~Shaw$^{\McGill,\Yale}$, 
E.~Shirokoff$^{\Caltech}$, 
 J.~S.~Spilker$^{\Arizona}$,
B.~Stalder$^{\CfA}$, 
Z.~Staniszewski$^{\Caltech,\CaseWestern}$, 
A.~A.~Stark$^{\CfA}$, 
K.~Story$^{\KICPChicago,\PhysicsUChicago}$, 
K.~Vanderlinde$^{\McGill}$, 
 N.~Welikala$^{\IAS}$,
R.~Williamson$^{\KICPChicago,\AAUChicago}$
}

\begin{document}

\maketitle

\begin{affiliations}
\item California Institute of Technology, 1200 E. California Blvd., Pasadena, CA 91125, USA
\item Steward Observatory, University of Arizona, 933 North Cherry Avenue, Tucson, AZ 85721, USA
\item Department of Physics and Atmospheric Science, Dalhousie University, Halifax, NS B3H 3J5 Canada
\item Institute of Astronomy, University of Cambridge, Madingley Road, Cambridge CB3 0HA, UK
\item European Southern Observatory, Karl-Schwarzschild Strasse, D-85748 Garching bei M\"unchen, Germany
\item Department of Physics, McGill University, 3600 Rue University, Montreal, Quebec H3A 2T8, Canada
\item Max-Planck-Institut f\"{u}r Radioastronomie, Auf dem H\"{u}gel 69 D-53121 Bonn, Germany
\item University of Pennsylvania, 209 South 33rd Street, Philadelphia, PA 19104, USA
\item University of Chicago, 5640 South Ellis Avenue, Chicago, IL 60637, USA
\item Harvard-Smithsonian Center for Astrophysics, 60 Garden Street, Cambridge, MA 02138, USA
\item Department of Physics, Harvard University, 17 Oxford Street, Cambridge, MA 02138, USA
\item Kavli Institute for Cosmological Physics, University of Chicago, 5640 South Ellis Avenue, Chicago, IL 60637, USA
\item Enrico Fermi Institute, University of Chicago, 5640 South Ellis Avenue, Chicago, IL 60637, USA
\item Department of Physics, University of Chicago, 5640 South Ellis Avenue, Chicago, IL 60637, USA
\item Jet Propulsion Laboratory, 4800 Oak Grove Drive, Pasadena, CA 91109, USA
\item Department of Physics and Astronomy, University of Missouri, 5110 Rockhill Road, Kansas City, MO 64110, USA
\item Department of Astronomy and Astrophysics, University of Chicago, 5640 South Ellis Avenue, Chicago, IL 60637, USA
\item Argonne National Laboratory, 9700 S. Cass Avenue, Argonne, IL, USA 60439, USA
\item Department of Physics,  University of California, One Shields Avenue, Davis, CA 95616, USA
\item National Radio Astronomy Observatory, 520 Edgemont Road, Charlottesville, VA 22903, USA
\item Department of Physics, University of California, Berkeley, CA 94720, USA
\item Department of Astronomy, University of Florida, Gainesville, FL 32611, USA
\item Department of Physics and Astronomy, University College London, Gower Street, London WC1E 6BT, UK
\item Department of Astrophysical and Planetary Sciences and Department of Physics, University of Colorado, Boulder, CO 80309, USA
\item Physics Division, Lawrence Berkeley National Laboratory, Berkeley, CA 94720, USA
\item Department of Physics and Astronomy, University of California, Los Angeles, CA 90095-1547, USA
\item Australia Telescope National Facility, CSIRO, Epping, NSW 1710, Australia
\item Department of Physics, University of Michigan, 450 Church Street, Ann Arbor, MI, 48109, USA
\item Physics Department, Center for Education and Research in Cosmology  and Astrophysics,  Case Western Reserve University, Cleveland, OH 44106, USA
\item Department of Astronomy, University of Michigan, 500 Church Street, Ann Arbor, MI, 48109, USA
\item Observatories of the Carnegie Institution for Science, 813 Santa Barbara Street, Pasadena, CA 91101, USA
%\item Physics Department, University of Minnesota, 116 Church Street S.E., Minneapolis, MN 55455, USA
\item Space Telescope Science Institute, 3700 San Martin Dr., Baltimore, MD 21218, USA
\item Liberal Arts Department, School of the Art Institute of Chicago,  112 S Michigan Ave, Chicago, IL 60603, USA
\item Department of Physics, Yale University, P.O. Box 208210, New Haven, CT 06520-8120, USA
\item Institut d'Astrophysique Spatiale, B\^atiment 121, Universit\'e Paris-Sud XI \& CNRS, 91405 Orsay Cedex, France
\end{affiliations}

\begin{abstract}

%%%%%%%%%%%%%%%%
% 1 Intro / abstract
%%%%%%%%%%%%%%%%

In the past decade, our understanding of galaxy  evolution has  been revolutionized by the discovery that luminous, dusty, 
starburst galaxies were 1,000 times more abundant in the early Universe than at present\cite{lagache05, chapman05}.
It has, however, been difficult to measure the complete redshift distribution of these objects, especially at the highest redshifts (${\bf z  > 4}$).
Here we report a redshift survey at a wavelength of three millimeters, targeting carbon monoxide line emission from the star-forming 
molecular gas in the direction of extraordinarily bright millimetre-wave-selected sources. 
High-resolution imaging  demonstrates that these sources are strongly gravitationally lensed by foreground galaxies.
We detect spectral lines in 23 out of 26 sources and multiple lines in 12 of those 23 sources, from which we obtain robust, unambiguous redshifts. 
At least 10 of the sources are found to lie at ${\bf z>4}$, indicating that the fraction of dusty starburst galaxies at high redshifts is greater than previously thought.  
Models of lens geometries in the sample indicate that the background objects are 
ultra-luminous infrared galaxies, powered by extreme bursts of star formation.

\end{abstract}

%%%%%%%%%%%%%%%%
% 2 description of selection and observations
%%%%%%%%%%%%%%%%

We constructed a catalog of high-redshift ($z>1$) galaxy candidates from the first 1,300\ square degrees of the South Pole Telescope (SPT)\cite{carlstrom11} survey 
by selecting sources with dust-like spectral indexes in the 1.4 and 2.0~mm SPT bands\cite{vieira10}. 
A remarkable aspect of selecting sources based on their flux at millimetre wavelengths is the so-called ``negative $k$-correction''\cite{blain93}, whereby
cosmological dimming is compensated by the steeply rising dust spectrum as the source redshift increases. 
As a result, a millimetre-selected sample should draw from the redshift distribution of
dusty starburst galaxies with little bias over the entire redshift range in which they are expected to exist. 
To isolate the high-redshift, dusty-spectrum galaxy population, sources with counterparts in the 
IRAS Faint Source Catalog\cite{moshir92} (typically $z<0.03$) were removed, and those with counterparts in the 
843~MHz Sydney University Molonglo Sky Survey\cite{bock98} were removed to exclude sources with strong synchrotron emission (e.g. flat-spectrum radio quasars) that 
may have passed the spectral index cut. A sample of 
47 sources with 1.4~mm flux density $>$ 20~mJy and accurate positions
were selected for high-resolution imaging with the Atacama Large Millimeter/sub-millimeter Array (ALMA).
The ALMA spectroscopic observations targeted a sample of 26 sources, all but two of which are in the imaging sample (see Supplementary Information). 
These objects are among the brightest dusty-spectrum sources in the $z>0.1$ extragalactic sky at millimetre wavelengths.

%%%%%%%%%%%%%%%%
% 3 imaging and lensing
%%%%%%%%%%%%%%%%

Gravitationally lensed sources are expected to predominate in samples of the very brightest dusty galaxies because of the 
rarity of unlensed dusty, starburst galaxies at these flux levels\cite{blain96, negrello07, hezaveh11}.  
Massive elliptical galaxies, acting as lenses, will have Einstein radii as large as $2\arcsec$ and may magnify background galaxies by factors 
of 10 or more.
To confirm the lensing hypothesis and determine magnifications, we imaged 47 SPT sources with ALMA at 870 \micron\ in two array configurations, which provide angular resolutions of  1.5\arcsec\ and 0.5\arcsec\ (FWHM).  
A sample of these objects with IR imaging, spectroscopic redshifts, and resolved structure is shown in Figure~1. 
Integration times of only one minute per source
are adequate to show that the sources are resolved into arcs or Einstein rings -- hallmarks of gravitational lensing -- in most sources. 
For all sources for which we have IR and submillimetre imaging, as well as spectroscopic redshifts,
the emission detected by ALMA coincides with massive foreground galaxies or galaxy groups/clusters, but is spatially distinct and at drastically different redshifts  
(see Figures~2 and S.3).
Using a modeling procedure that treats the interferometer data in their native measurement space, rather than through reconstructed sky images,
to simultaneously determine the source/lens configuration and correct for antenna-based phase errors\cite{hezaveh12b}, 
we are able to determine magnifications and derive intrinsic luminosities for our sources.
Complete models of four lenses\cite{hezaveh12b}, as well as preliminary models of eight more, indicate lensing 
magnifications between 4 and 22. After correcting for the magnification, these sources are
extremely luminous -- more than $10^{12} L_\odot$ and sometimes $>10^{13} L_\odot$ -- implying star formation rates in excess of $500~M_\odot $yr$^{-1}$.

%%%%%%%%%%%%%%%%
% 4 redshifts
%%%%%%%%%%%%%%%%

Obtaining spectroscopic redshifts for high-redshift, dusty starburst galaxies has been notoriously difficult. 
To date, most spectroscopic redshift measurements have come from 
the rest-frame ultraviolet and optical wavebands after multi-wavelength counterpart
identification\cite{ivison02, chapman05, coppin09}. 
These observations are difficult owing to the extinction of the UV light by the dust itself, the cosmological
dimming, and the ambiguity in the association of the dust emission with multiple sources of optical emission visible in deep observations. % with the source of the dust emission. 
A much more direct method to determine redshifts of %the spectroscopic redshift of
starburst galaxies, particularly at high redshift,  is through observations of molecular emission associated with their 
dusty star forming regions. The millimetre and submillimetre transitions of molecular carbon monoxide (CO) 
and neutral carbon (\ci) are well-suited for this purpose\cite{walter12}. 
These emission lines are a major source of cooling for the warm
molecular gas fueling the star formation, and can thus be related
unambiguously to the submillimetre continuum source\cite{solomon05}. 
Until recently, bandwidth and
sensitivity limitations made this approach time-intensive. 
The combination of ALMA -- even with its restricted early science capabilities and only 16 antennas -- and a unique sample of extraordinarily bright millimetre sources has changed this situation dramatically, allowing us to undertake a sensitive, systematic search for molecular and atomic lines across broad swaths of redshift space at $z>1$.

%%%%%%%%%%%%%%%%
% 5 n(z)
%%%%%%%%%%%%%%%%

We conducted a redshift search in the 3\,mm atmospheric transmission window with ALMA
using five spectral tunings of the ALMA receivers to cover $84.2-114.9$~GHz.  For $z>1$ at least 
one CO line will fall in this frequency range, except for a small redshift
``desert'' ($1.74<z<2.00$).  For sources at $z>3$, multiple transitions (such as
CO($J_{\rm up}>3$) and \cone) are redshifted into the observing band,
allowing for an unambiguous redshift determination. We find one or more 
spectral features in 23 of 26 SPT-selected sources. % yielding a $\sim90\%$ detection rate. 
The detections comprise 44 emission line features which we identify as 
redshifted emission from molecular transitions of $^{12}$CO, $^{13}$CO, H$_2$O and H$_2$O$^+$, and a \ci\ fine structure line.
The spectra of all sources are shown in Figure~2. 
For 18 of the sources we are able to infer unique redshift solutions, either from ALMA data alone (12), 
or with the addition of data from the Very Large Telescope and/or the Atacama Pathfinder Experiment telescope (6). 
With the 10  $z>4$ objects discovered here, we have more than doubled the number of 
spectroscopically confirmed, ultra-luminous galaxies discovered at $z>4$ in millimetre/submillimetre surveys in the literature 
(of which just nine have been reported previously\cite{capak08, daddi09a,daddi09b,  coppin09,  riechers10, cox11, combes12, walter12}).
Two sources are at $z=5.7$, placing them among
the most distant ultra-luminous starburst galaxies known.
%These results, incorporating less than 25\% of the entire SPT sample,  demonstrate the existence of dusty starburst galaxies at higher redshifts than were previously known and highlights the copious amounts of star-forming gas at these epochs. {\bf ** this sentence needs work ** }

%%%%%%%%%%%%%%%%
% 6 discussion of possible biases to redshifts
%%%%%%%%%%%%%%%%

The SPT dusty galaxy redshift sample comprises 28 sources, as we include an additional two SPT sources with spectroscopic 
redshifts\cite{greve12} that would have been included in the ALMA program had their redshifts not already been determined. 
Of the 26 ALMA targets, three lack a spectral line feature 
in the ALMA band. We tentatively and conservatively place these at $z=1.85$, in the middle of the $z=1.74-2.00$ redshift desert, though it is also possible that they are located at very high redshift or have anomalously faint CO lines. 
For the five sources for which only a single emission line is found, only two or three redshifts are possible (corresponding 
to two choices of CO transition) after excluding redshift choices for which the implied dust temperature, derived from our 
extensive millimetre/submillimetre photometric coverage (provided by ALMA 3\,mm, SPT 2\,\&\,1\,mm,
APEX/LABOCA 870~\micron\ and \textit{Herschel}/SPIRE 500, 350, 250~\micron\ observations\cite{weiss12}), is inconsistent with the 
range seen in other luminous galaxies\cite{greve12}.
%and 
%making reasonable assumptions about the dust temperature\cite{greve12}. 
For these sources we adopt the redshift corresponding to the dust temperature closest to the median dust temperature in the unambiguous spectroscopic sample, as shown in Figure~2. 
%and we note that the spectroscopic completeness is greatest at $z>3.5$, where multiple CO transitions will fall into the ALMA band.
%If we assume the lower redshift solution for these sources, the median redshift of our sample is $\langle z\rangle=3.1$, while for the higher redshift solution the median redshift of our sample is $\langle z\rangle=3.6$.

The cumulative distribution function of all 
redshifts in this sample is shown in Figure~3.
The median redshift of our full sample is $z_\mathrm{med}=3.5$. 
The redshift distribution of SPT sources with mm spectroscopic redshifts  is in sharp contrast to that of radio-identified
starbursts with optical spectroscopic redshifts, which have a significantly lower median redshift of $z_\mathrm{med}=2.2$, and for which only 15-20\% of the population 
is expected to be at $z>3$~\cite{chapman05}. 
Part of this difference can be attributed to the high flux threshold of the original SPT selection, which effectively requires 
that the sources be gravitationally lensed. A much smaller total volume is lensed at $z<1$ than at higher redshift, and, as expected, we do not find any such 
sources in the SPT sample\cite{weiss12}.
However, if we only compare sources at $z>2$ (the lowest confirmed spectroscopic redshift in the SPT sample), the median redshift of the radio-identified sample is still significantly lower (2.6) than the SPT sample, and the probability that both samples are drawn from the same distribution is $<10^{-5}$ by the Kolmogorov-Smirnov test.
A recently published survey\cite{smolcic12} of millimetre-identified starbursts with optical counterparts determined from high resolution mm imaging and redshifts measured from optical spectroscopy or estimated from optical photometry Ê
found a median redshift Êof $z_\mathrm{med}=2.8$.
Again comparing the distribution of sources at $z>2$, the probability that these objects and the SPT-selected sources are drawn from the same parent distribution is 0.43, indicating rough consistency between our secure redshift determinations and the distribution estimated from the optical methods.
A full analysis of the molecular line detections, redshift determinations, residual selection effects, and a derivation of the intrinsic redshift distribution for the SPT sample is reported in a companion paper\cite{weiss12}.

%%%%%%%%%%%%%%%%
% 7 wrap up.
%%%%%%%%%%%%%%%%

These 26 sources represent less than 25\% of the recently completed SPT survey and catalog.
This newly discovered population of high-redshift starbursts will enrich our understanding of obscured star formation in the early 
universe. Existing semi-analytic hierarchical models of galaxy evolution\cite{baugh05,benson12b} have already had 
difficulties reconciling the number of $z\approx4$ systems inferred from previous observational studies\cite{coppin09,smolcic12}. 
The presence of two intensely starbursting galaxies at $z=5.7$, 1~Gyr after the Big Bang, in a sample of just 26 sources, demonstrates that significant 
reservoirs of dust and molecular gas had been assembled by the end of  the epoch of cosmic reionization. 
As the millimetre-brightest high-redshift starbursts in the sky, the present sample will be key targets for ALMA studies of 
star formation physics at high redshift. The gravitational lensing of these sources provides access to diagnostic information from molecular 
lines that would otherwise take hundreds of times longer to observe, and effective source-plane resolution several times higher than can 
otherwise be achieved.

%% Here is the endmatter stuff: Supplementary Info, etc.
%% Use \item's to separate, default label is "Acknowledgements"

\clearpage

%%%%%%%%%%%%%%%%
% Bibliography
%%%%%%%%%%%%%%%%

%\bibliographystyle{apj}
\bibliography{../bibtex/spt_smg}

\clearpage

\begin{addendum}

\item[Acknowledgements] 
The SPT is supported by the National Science Foundation, the Kavli Foundation and the Gordon and Betty 
Moore Foundation. 
ALMA is a partnership of ESO (representing its member states), NSF (USA) and NINS (Japan), together with NRC (Canada) and NSC and ASIAA (Taiwan), in cooperation with the Republic of Chile. 
The Joint ALMA Observatory is operated by ESO, AUI/NRAO and NAOJ.
The National Radio Astronomy Observatory is a facility of the National Science Foundation operated under cooperative agreement by Associated Universities, Inc. 
Partial support for this work was provided by NASA from the Space Telescope Science Institute. 
This work 
is based in part on observations made with {\it Herschel}, a European Space Agency Cornerstone Mission with significant participation 
by NASA.
Work at McGill is supported by NSERC, the CRC program, and CIfAR.

\item[Author Contributions] 
JDV and DPM wrote the text. SCC took and reduced optical images and spectroscopy. AW, CDB, and DPM analyzed the ALMA spectra. DPM, JSS and YH analyzed the ALMA imaging data. JDV reduced and analysed the \textit{Herschel} data.   
YH constructed the lens models. 
CDF reduced optical images. 
All other authors (listed alphabetically) have contributed as part of the South Pole Telescope collaboration, by either their involvement with the construction of the instrument, the initial discovery of the sources, or multi-wavelength follow-up, and/or contributions to the text. %All co-authors have read and commented on the text.

\item[Competing Interests] 
The authors declare that they have no competing financial interests.

 \item[Correspondence] Correspondence and requests for materials
should be addressed to J.D.V.~(email: vieira@caltech.edu).

\end{addendum}

\clearpage

%%%%%%%%%%%%%%%%
% PLOTS
%%%%%%%%%%%%%%%%

\begin{figure}[h] %  figure placement: here, top, bottom, or page
\begin{center}
\begin{tabular}{c} 
\includegraphics[width=17cm, trim=2.2cm 1.cm 0cm 0.7cm, clip=true]{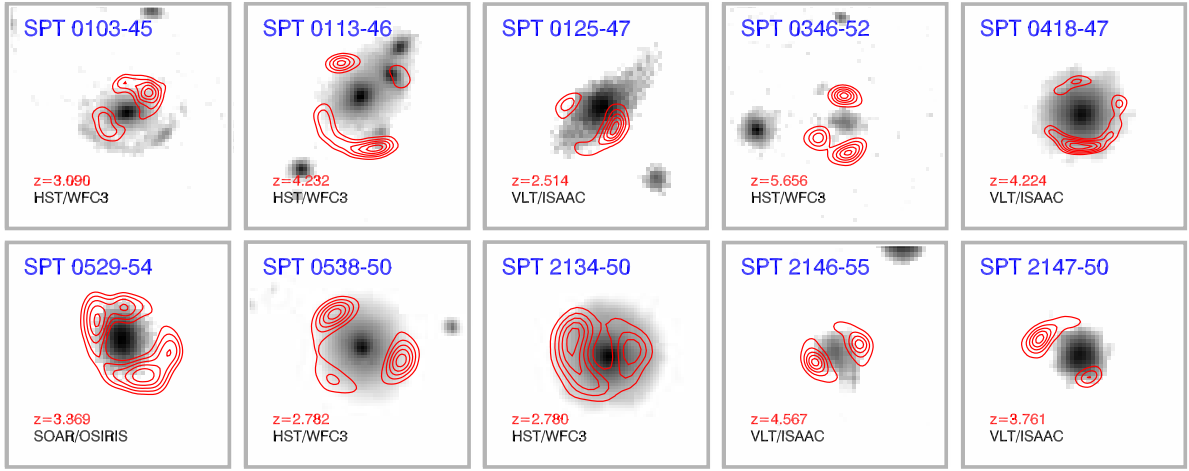} \\
\end{tabular}
\end{center}
\vspace{-2.2em}
\caption{\small \textbf{Figure 1:} Near-infrared (NIR) and ALMA submillimeter-wavelength images of SPT targets. 
Images are 8\arcsec$\times$8\arcsec. We show 10 sources for which we have
confirmed ALMA spectroscopic redshifts, deep NIR imaging, and well-resolved structure in the ALMA 870~\micron\ imaging. 
The greyscale images are NIR exposures from the \textit{Hubble Space Telescope} (co-added F160W and F110W filters), 
the Very Large Telescope ($K_s$), or the Southern Astrophysical Research Telescope ($K_s$), and trace the starlight from the foreground lensing galaxy. The NIR images are shown with logarithmic stretch.
 The red contours are ALMA 870 \micron\  imaging showing the background source structure, clearly indicative of strong lensing from galaxy-scale halos.  In all cases, the contours start at 5 $\sigma$ and are equally spaced up to 90\% of the peak significance, which ranges from 12 to 35.
Spectroscopic redshifts of the background sources are shown in red in each panel.
The ALMA exposures were approximately 2-minute integrations, roughly equally divided between the compact 
and extended array configurations. The resulting resolution is 0.5\arcsec. 
SPT~0103$-$45 shows a rare lensing configuration of one lens and two background sources at different redshifts, one visible 
with ALMA and one with HST.  SPT~0346$-$52, with a CO-derived redshift of $z$=5.656, is among the highest-redshift starbursts known. (See the Supplementary Information for more details.)
}
  \label{fig:fig1}
\end{figure}

\clearpage

\begin{figure}[h] %  figure placement: here, top, bottom, or page
\begin{center}
\begin{tabular}{c} 
\includegraphics[width=16cm, trim=0cm 0cm 0cm 0cm, clip=true]{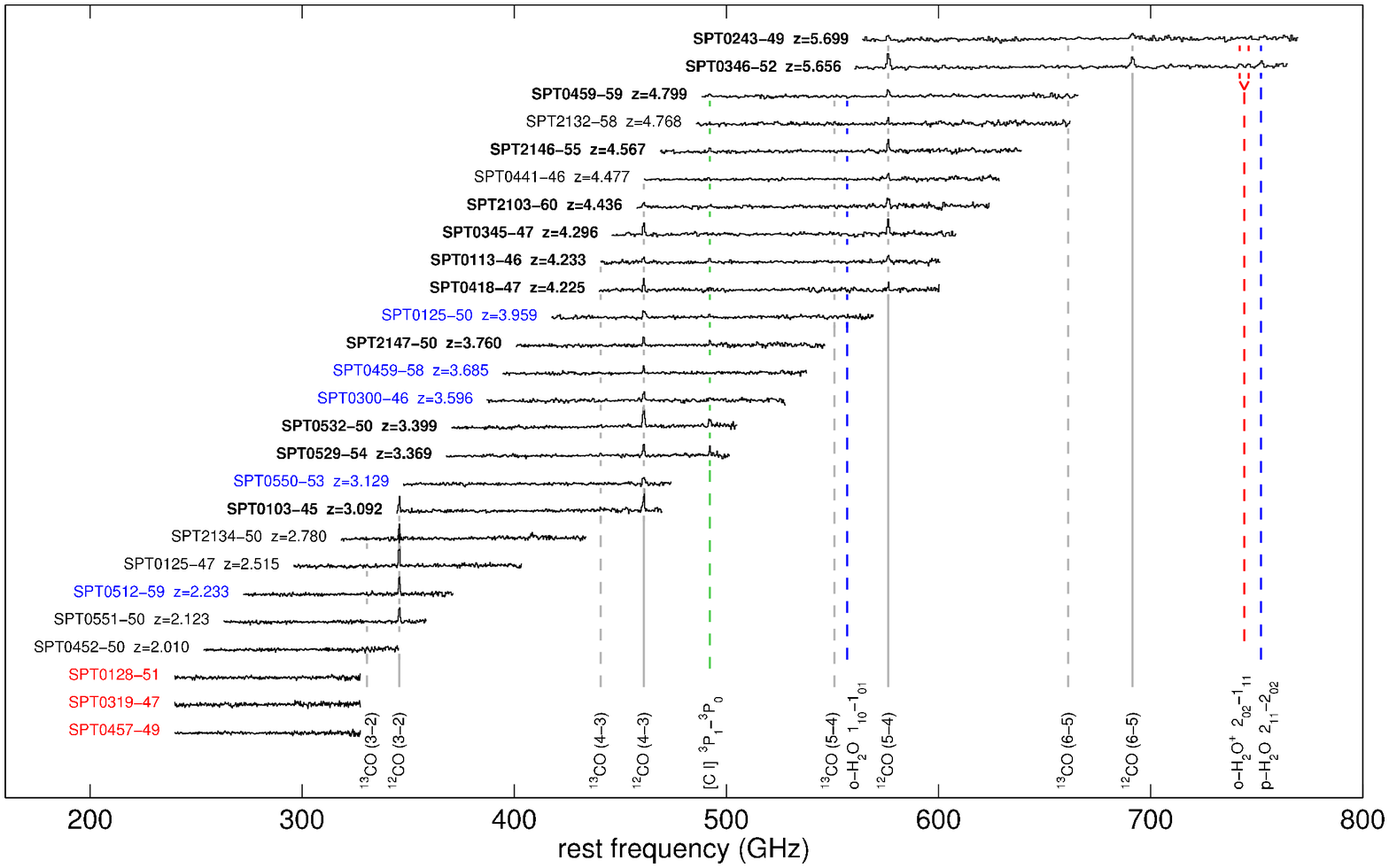} \\
\end{tabular}
\end{center}
\vspace{-2.2em}
\caption{\small \textbf{Figure 2:}  ALMA 3\,mm spectra of 26 SPT sources. The vertical axis is observed flux density in units of mJy, with 30~mJy offsets between sources for clarity. Spectra are continuum-subtracted. The strong CO lines are indicative of dust-enshrouded active star formation. 
   The spectra are labeled by source and redshift. Black labels indicate unambiguous redshifts (18), with the subset in {\bf bold} (12) having been derived from 
   the ALMA data alone. 
   Five sources labeled in blue (5) are plotted at the most likely redshift of multiple options, based on the dust temperature derived from extensive far-infrared photometry.
   Three sources with no lines detected are placed at $z=1.85$, in the middle of the redshift range for which we expect no strong lines, and labeled in red. 
Total integration times for each source were roughly ten minutes. The synthesized beam size ranges from 7\arcsec$\times$5\arcsec\ to 
5\arcsec$\times$3\arcsec\ over the frequency range of the search, which is inadequate to spatially 
resolve the velocity structure of the lensed sources.
}
  \label{fig:fig2}
\end{figure}

\clearpage

\begin{figure*}[ht]
\centering
\includegraphics[width=8.cm,angle=0]{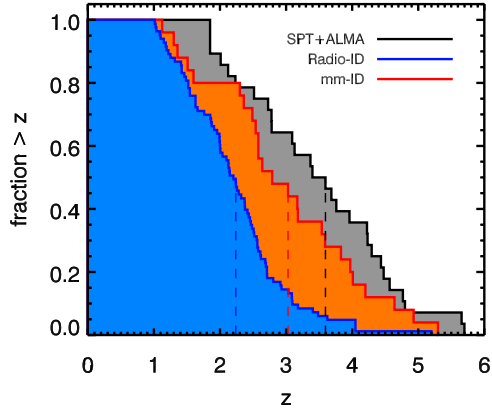}

\caption{ \small \textbf{Figure 3:} The cumulative redshift distribution of luminous, dusty starburst galaxies, as measured with different techniques. 
The SPT millimetre-selected sample, with redshifts directly determined from spectroscopic observations of the molecular 
gas in the galaxies, is shown in black. 
The existing samples of radio-identified starbursts\cite{chapman05, daddi09a, daddi09b, banerji11,walter12} with redshifts 
determined from rest-frame ultraviolet spectroscopy, are compiled in the blue distribution. 
The redshift distribution\cite{smolcic12} of millimetre-identified starburst galaxies in the COSMOS survey is shown in red/orange, 
though the majority of redshifts in this sample are derived from optical/IR photometry of the sources rather than spectroscopy, 
and therefore less certain.
Sources at $z<1$ were removed from the previous samples of starburst galaxies to better compare to the selection effect imposed on the SPT sample due to gravitational lensing. 
The distribution of redshifts for radio-identified sources is incompatible with the distribution for the sample presented in this work. 
This measurement demonstrates that the fraction of dusty starburst galaxies at high-redshift is greater than previously derived and that radio-identified samples were biased to lower redshift than the underlying population.
}
\label{fig:fig3}
\end{figure*}

\clearpage

%%%%%%%%%%%%%%%%
% Appendix
%%%%%%%%%%%%%%%%

\clearpage

\appendix
\pagestyle{empty}
\noindent{\bf Sample Selection}\\
The dusty-spectrum sources targeted for the ALMA observations described here were found in the SPT survey. 
The full survey comprises 2540 deg$^2$ of mapped sky, but constraints of data analysis and followup limited the area available 
for target selection at the time of the ALMA Cycle 0 deadline to 1300 deg$^2$. The initial SPT target list was extracted from 
the SPT maps according to the procedure described in a previous paper\cite{vieira10} and the selection outlined in the manuscript. 
Before observing these sources with 
ALMA, we required that they have followup observations with the LABOCA 870~\micron\  
camera on the Atacama Pathfinder Experiment telescope to improve the accuracy 
of their positions. At the time of the ALMA Cycle 0 proposal deadline, we had completed this followup for 
76 sources within 1300 deg$^2$ of the survey area. We selected 47 sources for imaging and 26 sources for 
spectroscopy with ALMA; 24 of the 26 spectroscopic sources were also in the imaging sample. 
The sample selection targeted sources with the highest SPT 1.4~mm fluxes, subject to the restrictions of the ALMA call for proposals.
The most important restriction was the requirement 
that sources be located within 15 degrees of each other on the sky. This should not affect the statistical properties of the sample, 
however, it merely prevented the observation of a complete set of SPT sources above a defined flux threshold.

\noindent{\bf Detected line features} \\
The detected CO and \ci\ line features are shown in Figure~\ref{fig:fig4}. Additional lines are detected in some spectra, 
including $^{13}$CO transitions in two sources. However, the detection of both $^{12}$CO and 
$^{13}$CO transitions in the same source does not break redshift degeneracies because both transitions are harmonically 
spaced; at a given frequency of detection, every pair of CO isotopic transitions of the same rotational level ($J$) will have the same 
observed spacing. Emission lines of H$_2$O and H$_2$O$^+$ are detected in the spectrum of SPT0346-52.

\noindent {\bf ALMA imaging} \\
Continuum images from the ALMA observations at both wavelengths are shown in Figure~\ref{fig:fig5} for all 49 sources observed 
with ALMA in one or both of the 3~mm redshift search or the 870~\micron\ imaging projects. 
The positional coincidence between the bands confirms that the redshifts are derived 
for the same objects that are seen to show structures indicative of gravitational lensing.
Nearly all sources are resolved at the 0.5\arcsec\ resolution of the 870~\micron\ data, most likely due to gravitational lensing. 
Exceptions may be due to lensing by groups/clusters, with image counterparts that are either faint or too widely separated to be detected in the small 
(18\arcsec) primary beam of the ALMA antennas at this wavelength, or because some of our objects have small image separations. Lensed dusty sources 
with similar image separations are already prominent in the literature, including APM~08279+5255\cite{egami00}, 
which has three images separated by $<$0.4\arcsec. Conclusive evidence of lensing in many objects, including the most compact, awaits 
IR imaging and spectroscopic redshifts for the sources (and any candidate lens galaxies). The 23 objects for which we have the most complete 
data (the 26 sources of the 3~mm spectroscopic sample, less three without detected lines, less two without IR imaging, augmented by 
the two sources\cite{greve12} for which we had prior redshifts) are shown in Figure~\ref{fig:figAll}.

\renewcommand{\thefigure}{S.\arabic{figure}}

\begin{figure*}[ht]
\centering
\includegraphics[width=13.cm,angle=0]{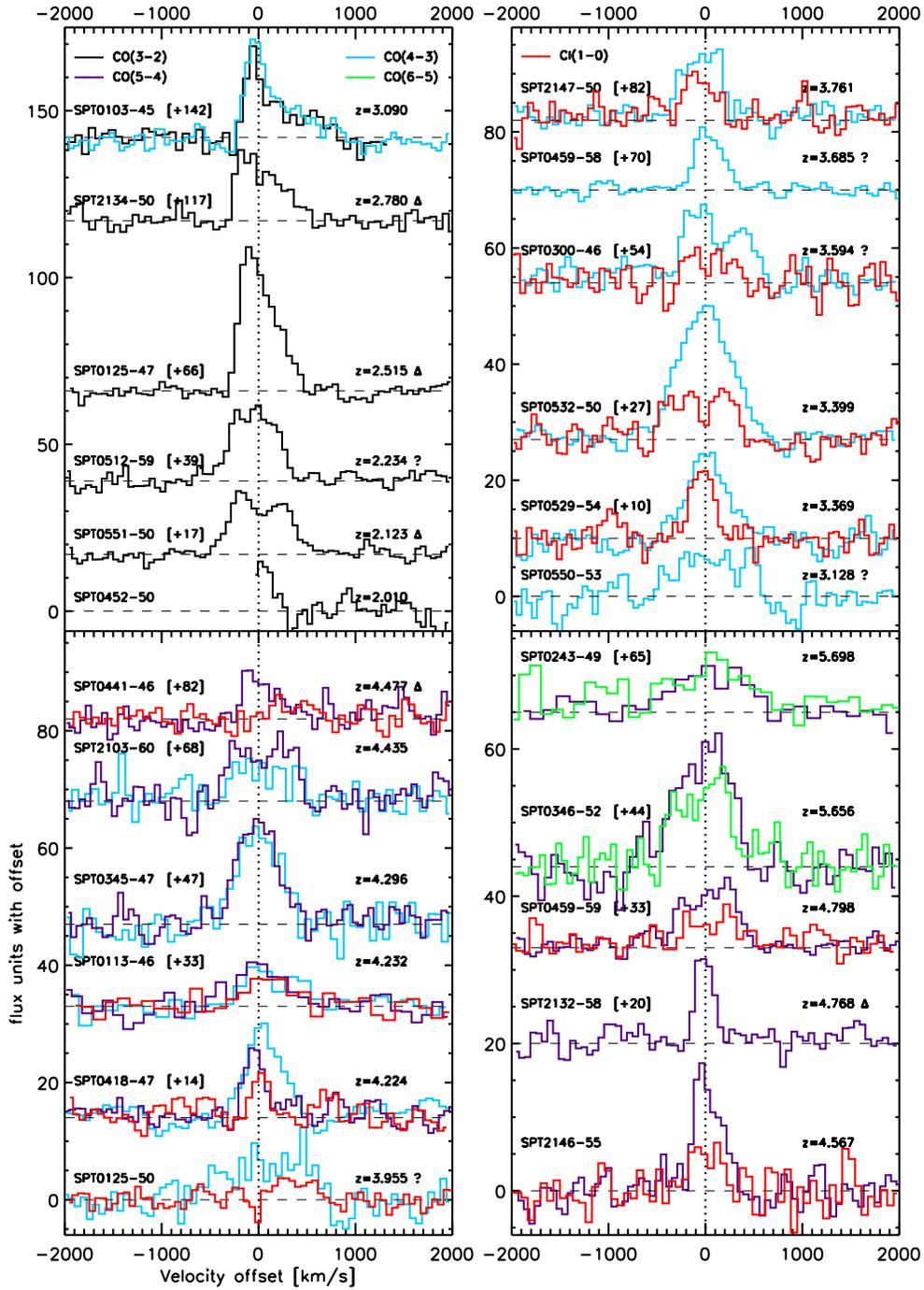}

\caption{The CO(3-2), CO(4-3), CO(5-6), CO(6-5) and \ci(1-0) emission lines observed with ALMA for 23 out of the 26 SPT sources, which were used to determine the source redshifts. The vertical axis is observed flux density, sources are offset from zero for clarity with the 
offsets specified in square brackets next to the source names. 
Redshifts marked with `$\Delta$' are confirmed using additional observations from other facilities\cite{weiss12}, 
while redshifts marked with `?' are uncertain and are shown at the most likely redshift. SPT0452-50 has a single line, but is determined 
to be at $z=2.010$ rather than $z=1.007$ because the implied dust temperature for this source would be far lower than in 
any other source (13~K) were it at the lower redshift\cite{weiss12}.}
\label{fig:fig4}
\end{figure*}

\begin{figure*}[ht]
\centering
\includegraphics[width=15.cm,trim=2.2cm 1.4cm 0.7cm 0.7cm, clip=true]{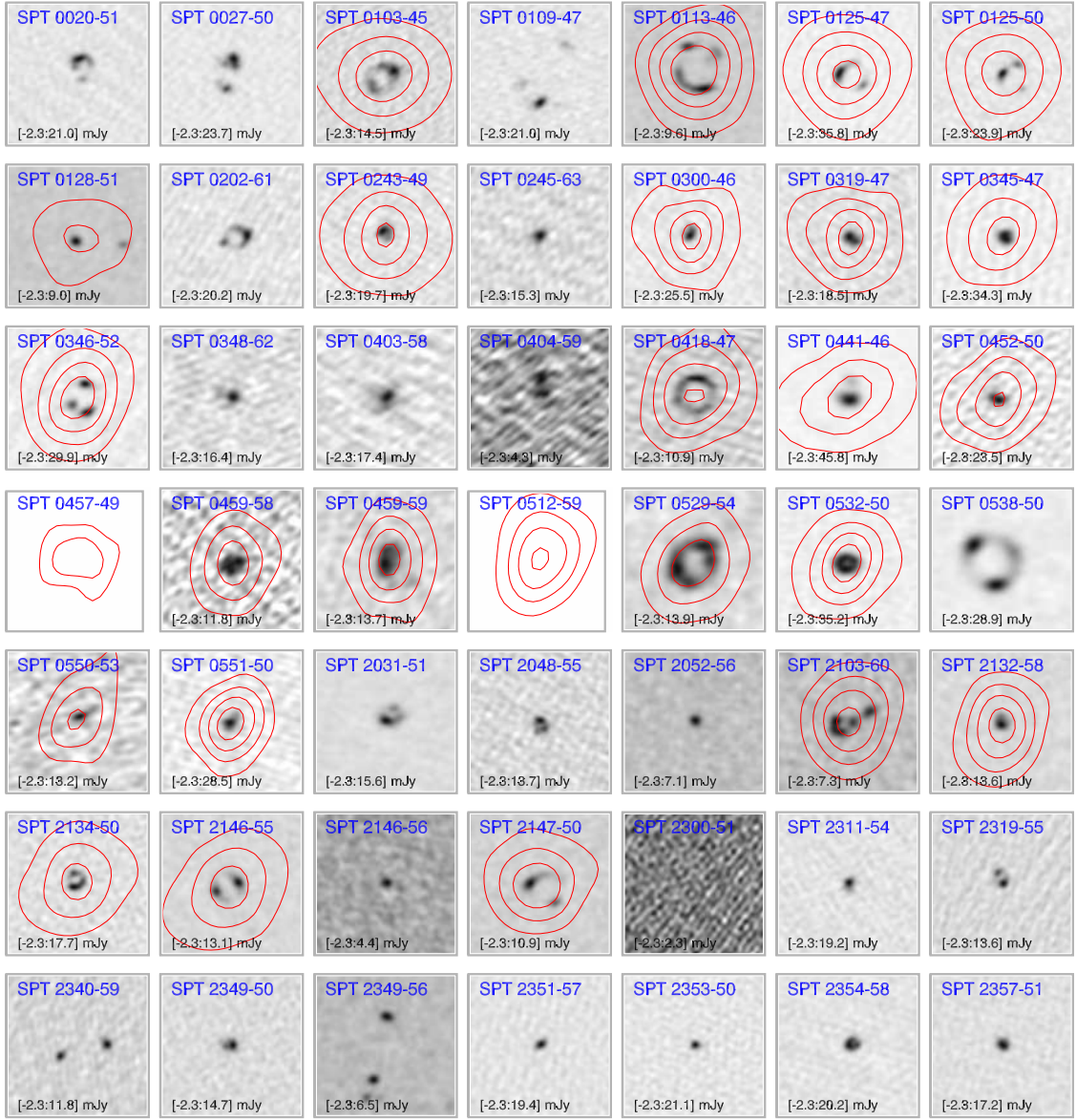}

\caption{Continuum images of the 49 sources observed at 3~mm and/or 870~\micron\ wavelength with ALMA. 
The 47 sources for which 870~\micron\ data were acquired are shown in greyscale, with the 3~mm images 
overlaid in red contours. Two sources from the redshift sample that lack 870~\micron\ data appear as red contours 
on a blank background. Images are 10\arcsec$\times$10\arcsec, the 870~\micron\ and 3~mm images have 
0.5\arcsec\ and 5\arcsec\ resolution, respectively. The correspondence between the positions at the two wavelengths
unambiguously links the lensing structure visible at 870~\micron\ to the 3~mm spectra. 
The 3~mm contours are plotted in units of 3$\sigma$, starting at 3$\sigma$ for sources at S/N$<$15, and 5$\sigma$ for sources at S/N$>$15, 
except SPT~0457-49, where the contours are 3 and 4$\sigma$. 
The grey scale stretch of each image is indicated in the lower left hand side of each panel and is roughly from $-1\sigma$ to the peak value. 
}
\label{fig:fig5}
\end{figure*}

\begin{figure*}[ht]
\centering
\includegraphics[width=16.5cm, trim=2.2cm 1.4cm 0.7cm 0.6cm, clip=true]{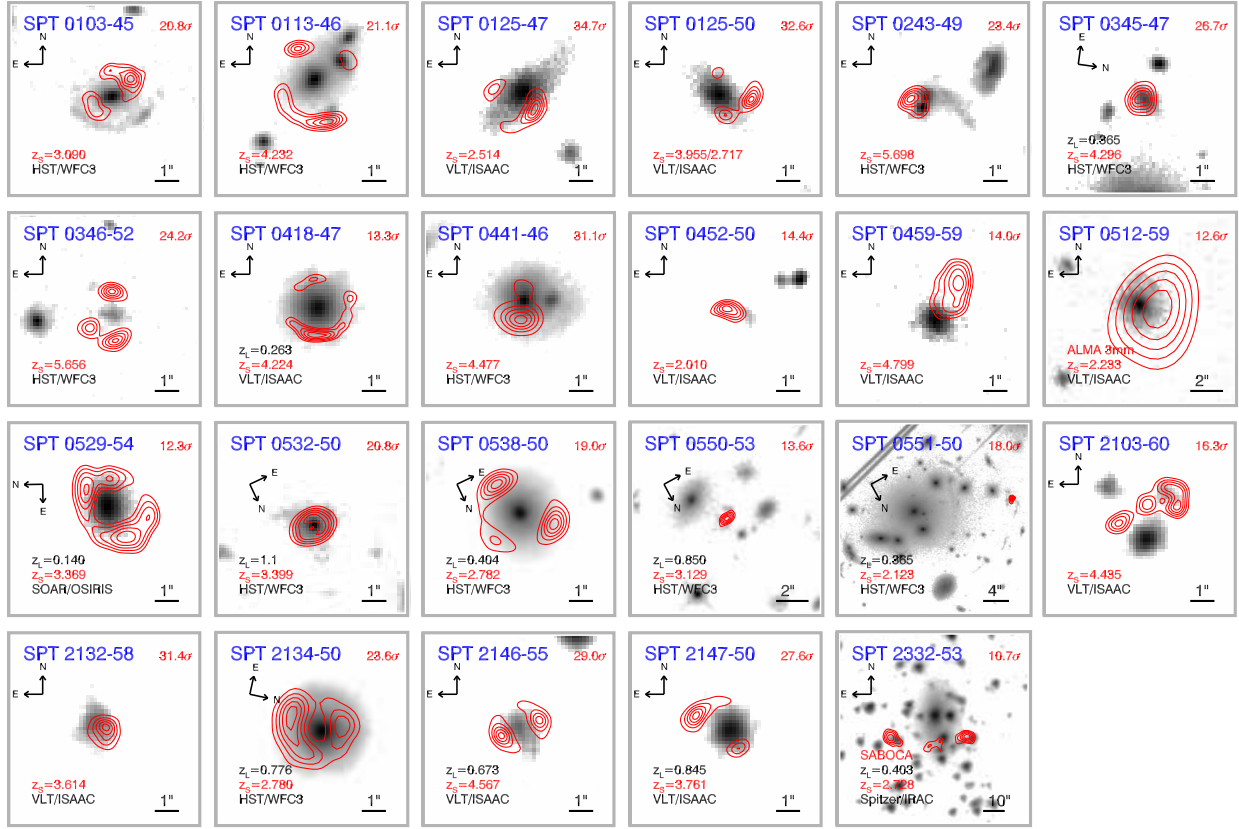} \\
\caption{Images of the full set of 23 sources for which we have ALMA 870~\micron\, 3~mm, or SABOCA 350~\micron\ imaging, deep NIR imaging, and a 
redshift for the background galaxy (including ambiguous redshifts). 
Except for SPT~0512-59, ALMA 870~\micron\ emission is represented with 5 red contours, spaced linearly from five times the image noise to 90\% of the peak 
signal to noise, specified in the upper right of each panel. 
For SPT~0512-59, which lacks ALMA 870~\micron\ data, we show the ALMA 3~mm continuum contours. 
For SPT~2332-53, which lacks ALMA 870~\micron\ data, we show the APEX/SABOCA 350~\micron\ continuum contours. 
The redshift of the background source ($z_{\rm S}$) is specified in red. 
Greyscale images are near-infrared exposures from the \textit{Hubble Space Telescope} (co-added F160W and F110W filters), 
the Very Large Telescope ($K_s$), the Southern Astrophysical Research Telescope ($K_s$), or the 
{\it Spitzer Space Telescope} (3.6~\micron) and trace the starlight from the foreground lensing galaxy. 
The images are shown with logarithmic stretch. When known, the redshift of the foreground galaxy ($z_{\rm L}$) is specified in black.
In nearly every case, there is a coincidence of the millimetre/submillimetre emission, determined by the 
redshift search data to arise at high redshift, with a lower redshift galaxy, a galaxy group, or a cluster. This is precisely the expectation 
for gravitationally lensed galaxies. Three cluster lenses are apparent, SPT~0550-53, SPT~0551-50, and SPT~2332-53, 
with two other systems lensed by compact groups (SPT~0113-46, SPT~2103-60). 
}
\label{fig:figAll}
\end{figure*}

%%
%% TABLES
%%
%% If there are any tables, put them here.
%%

\end{document}